\documentclass[twocolumn,showpacs,preprintnumbers,amsmath,amssymb]{revtex4}
\usepackage{graphicx}
\usepackage{textcomp}

\begin{document}
\draft
\title{Periodic wave packet reconstruction in truncated tight-binding lattices}
  \normalsize

\author{Stefano Longhi\footnote{Author's email address:
longhi@fisi.polimi.it}}
\address{Dipartimento di Fisica, Politecnico di Milano, Piazza L. da Vinci 32, I-20133 Milano, Italy}


%
\bigskip
\begin{abstract}
\noindent A class of truncated tight-binding Hermitian and
non-Hermitian lattices with commensurate energy levels, showing
periodic reconstruction of the wave packet, is presented. Examples
include exact Bloch oscillations on a finite lattice, periodic wave
packet dynamics in non-Hermtian lattices with a complex linear
site-energy gradient, and self-imaging in lattices with commensurate
energy levels quadratically-varying with the quantum number.
\end{abstract}


\maketitle {\em Introduction.} Tight-binding lattice models provide
a simple tool to investigate the transport properties of different
physical systems, such as electronic transport in semiconductor
superlattices or coupled quantum dots, photonic transport in
waveguide arrays or coupled optical cavities, and transport of cold
atoms in optical lattices. In certain conditions, the dynamics of a
wave packet in the lattice turns out to be periodic. Such a special
regime is attained, for example, in an infinitely-extended periodic
lattice with a superimposed dc field, leading to the formation of an
equally-spaced Wannier-Stark ladder energy spectrum and to the
appearance of Bloch oscillations (BOs) \cite{Gluck02}. Another
important case is that of an infinitely-extended periodic lattice
with an applied ac field, for which quasi-energy band collapse
leading to dynamic localization can be found at special ratios
between amplitude and frequency of the ac modulation
\cite{Dunlap86,Holthaus92}. Such periodic dynamical behaviors
realize wave self-imaging, i.e. a periodic reconstruction of the
initial wave packet distribution. Self-imaging phenomena in
tight-binding lattices have been observed in different physical
systems (see, for instance, \cite{BO1,BO2,DL1,DL11,DL2}). Even in an
infinite periodic lattice without any external applied field
periodic self-imaging can be observed for special periodic initial
wave distributions owing to the discrete analogue of the Talbot
effect \cite{Talbot}. Unfortunately, lattice truncation, defects and
edge effects generally destroy the periodic wave packet dynamics
because of incommensurate frequencies of the eigenstates. As a
consequence, the system never revives fully to its initial state,
and phenomena like wave packet collapse and revivals are generally
found \cite{Longhi08}. Recently, a few examples of lattice
engineering that enables to restore exact periodic wave packet
dynamics have been proposed
\cite{Lambropoulos,Gordon,SzameitAPL,Longhi09,Longhi10}, including
harmonic oscillations in a finite lattice with engineered hopping
rates \cite{Lambropoulos,Gordon} and Bloch oscillations on a
semi-infinite lattice with linearly-increasing hopping rates
\cite{Longhi09}. In this Report we introduce a more general class of
truncated tight-binding lattices -not necessarily Hermitian- that
sustain a periodic dynamics of the wave packet and that include the
Hermitian lattices of Refs.\cite{Lambropoulos,Gordon,Longhi09} as
special cases. In particular, we prove the existence of Bloch
oscillations on engineered finite lattices, either Hermitian or
non-Hermitian, and show self-imaging phenomena in a novel class of
finite lattices with commensurate energy levels
that depend quadratically on the quantum number.\\
\\
{\em Lattice synthesis.} Let us consider  a tight-binding
one-dimensional lattice with site energies $V_n$ and hopping
amplitude $\kappa_n$ between adjacent sites $|n \rangle$ and $|n+1
\rangle$. In the nearest-neighboring approximation, indicating by
$\psi(n)$ the occupation amplitude at lattice site $|n \rangle$, the
energies $E$ of the lattice Hamiltonian $\mathcal{H}$ are found as
eigenvalues of the second-order difference equation
\begin{equation}
\mathcal{H} \psi(n) \equiv \kappa_{n-1} \psi(n-1)+\kappa_{n}
\psi(n+1)+V_n \psi(n)=E \psi(n).
\end{equation}
Lattice truncation is realized by assuming $\kappa_{-1}=0$ for a
semi-infinite lattice comprising the sites $|0 \rangle$, $|1
\rangle$, $|2 \rangle$ ...., or $\kappa_{-1}=\kappa_{N}=0$ for a
finite lattice comprising the $(N+1)$ sites $|0 \rangle$, $|1
\rangle$, $|2 \rangle$, ... $|N \rangle$. Note that the Hamiltonian
$\mathcal{H}$ is Hermitian provided that $V_n$ and $\kappa_n$ are
real-valued, the emergence of complex values for either hopping
rates or site energies being the signature of a non-Hermitian
lattice. Our aim is to judiciously engineer the hopping amplitudes
$\kappa_n$ and site energies $V_n$ in such a way that the lattice
energies $E$ form a set of commensurate numbers, thus ensuring a
periodic dynamics of an arbitrary wave packet $\psi(n,t)$ that
evolves according to the time-dependent Schr\"{o}dinger equation $ i
(\partial \psi/ \partial t)=\mathcal{H} \psi$. Examples of such a
lattice engineering, corresponding to a set of equally-spaced
discrete energy levels, have been previously presented in
Refs.\cite{Lambropoulos,Gordon,Longhi09}. Here we show that such
previous examples belong to a more general class of lattices with
commensurate energies, which can be synthesized using the
factorization method \cite{susy1}, or the so-called supersymmetric
quantum mechanics \cite{susy2}, adapted to 'discrete' quantum
mechanics and difference equations (see, for instance, \cite{calo}).
To this aim, let us assume that the hopping amplitudes $\kappa_n$
and site energies $V_n$ can be derived from two functions $F_1(n)$
and $F_2(n)$ according to
\begin{equation}
\kappa_n=\sqrt{F_1(n)F_2(n+1)} \; , \; V_n=-[F_1(n)+F_2(n)].
\end{equation}
In this case, the Hamiltonian $\mathcal{H}$ can be factorized as
$\mathcal{H}=\mathcal{A}\mathcal{B}$, where \cite{calo}
\begin{eqnarray}
\mathcal{A} & = & \sqrt{F_1(n)} \exp \left( \frac{1}{2}
\frac{\partial}{\partial n} \right)- \sqrt{F_2(n)} \exp \left(
-\frac{1}{2} \frac{\partial}{\partial n} \right)\;\;\;\;\; \\
\mathcal{B} & = & \exp \left( \frac{1}{2} \frac{\partial}{\partial
n} \right) \sqrt{F_2(n)} - \exp \left( -\frac{1}{2}
\frac{\partial}{\partial n} \right) \sqrt{F_1(n)}. \;\;\;\;\;
\end{eqnarray}
To realize a finite lattice, with nonvanishing amplitudes $\psi(n)$
at the $(N+1)$ lattice sites $|0\rangle$, $|1\rangle$, ...,
$|N\rangle$, we require $F_1(N)=0$ and $F_2(0)=0$. For a
semi-infinite lattice, we require $F_2(0)=0$ solely. Let us then
indicate by $\psi_0(n)$ the solution to the difference equation
$\mathcal{B} \psi_0(n)=0$, i.e. $\sqrt{F_2(n+1)}
\psi_0(n+1)=\sqrt{F_1(n)} \psi_0(n)$. It then follows that
$\mathcal{H} \psi_0=0$, i.e. $E=0$ is an eigenvalue of $\mathcal{H}$
\cite{note1}. To find the other eigenvalues of $\mathcal{H}$, let us
search for a solution to Eq.(1) in the form $\psi(n)=\psi_0(n)Q(n)$;
it then readily follows that $Q(n)$ satisfies the second-order
difference equation
\begin{equation}
F_1(n) Q(n+1)+F_2(n)Q(n-1)=[F_1(n)+F_2(n)+E] Q(n).
\end{equation}
Assuming for $F_1(n)$ and $F_2(n)$ a linear or quadratic functions
of $n$, Eq.(5) can be solved in an exact way by assuming for $Q(n)$
a polynomial of degree $M$, i.e. $Q(n)=\sum_{l=0}^{M} a_l n^l$.
Substitution of such an expression of $Q(n)$ into Eq.(5) yields the
following relation
\begin{eqnarray}
\sum_{\rho=0}^{M-1} \left( \sum_{k=\rho+1}^{M} a_k \left(
\begin{array}{c}
k \\ \rho
\end{array}
\right)  \left [ F_1(n)+(-1)^{k-\rho}F_2(n) \right] \right) n^{\rho}
\nonumber \\
=E \sum_{\rho=0}^{M}a_{\rho}n^{\rho} \;\;\;\; \;\;\;\; \;\;\;\;
\;\;\;\; \;\;\;\;
\end{eqnarray}
which determines the eigenvalues $E$ and the coefficients $a_{\rho}$
of the polynomial $Q(n)$ after comparison of the terms of the same
power $n^{\rho}$ on the left- and right-hand-sides in Eq.(6). In
particular, the eigenvalue $E$ is found by comparison of the
highest-order power terms $n^M$. It is worth discussing in a
separate way the cases of linear and quadratic forms for $F_1(n)$
and $F_2(n)$, which yield a different dependence of the energy
eigenvalues on the quantum number $M$.\\
{\it Lattices with equally-spaced energy levels: Bloch oscillations
on a finite lattice}. Let us consider a finite lattice comprising
the sites $n=0,1,2,...,N$ and assume $F_1(n)=\alpha(n-N)$ and
$F_2(n)=\gamma n$, with $\alpha$ and $\gamma$ arbitrary parameters
with $\alpha \neq \gamma$. The hopping amplitudes and site energies
of this lattice are thus given by [see Eq.(2)]
\begin{equation}
\kappa_n=\sqrt{\alpha \gamma (n-N)(n+1)} \; , \; \; V_n=\alpha
N-(\alpha+\gamma)n.
\end{equation}
Note that the lattice Hamiltonian is Hermitian for $\alpha
\gamma<0$, whereas it is not Hermitian for $\alpha \gamma>0$ or for
complex-values for $\alpha$ and/or $\gamma$. Substitution of such
expressions of $F_1(n)$ and $F_2(n)$ into Eq.(6) and by comparing
the coefficient of the highest-degree term ($n^M$), yields the
following simple expression for the eigenvalues
\begin{equation}
E=M(\alpha-\gamma),
\end{equation}
\begin{figure}[htbp]
  \includegraphics[width=78mm]{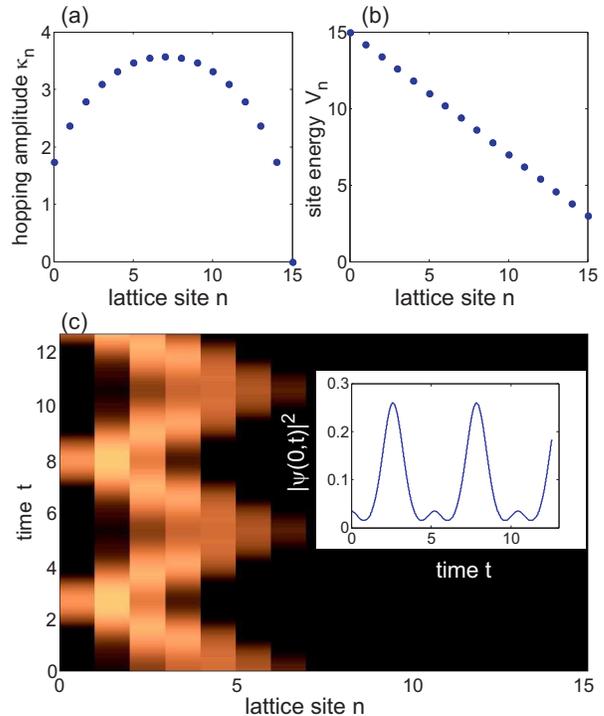}\\
   \caption{(color online) Exact Bloch oscillations in an Hermitian truncated lattice. (a) coupling
   amplitudes $\kappa_n$, and (b) lattice site energies $V_n$ as given by Eq.(7) for
$N=15$, $\alpha=1$ and $\gamma=-0.2$. (c) Numerically-computed
temporal evolution of the site occupation probabilities
$|\psi(n,t)|^2$ in the $(n,t)$ plane for an input  Gaussian wave
occupation amplitudes $\psi(n,0)=\exp\{-[(n-N/4)/4]^2 \}$. The inset
in (c) shows the detailed temporal evolution of the occupation
probability $|\psi(0,t)|^2$ of the left-edge ($n=0$) lattice site.}
\end{figure}
\begin{figure}[htbp]
  \includegraphics[width=78mm]{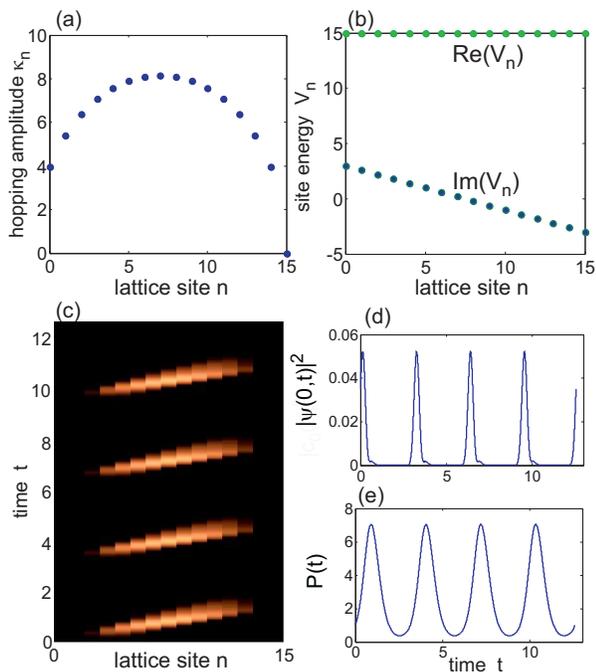}\\
   \caption{(color online) Bloch oscillations in a non-Hermitian truncated lattice. (a) coupling
   amplitudes $\kappa_n$, and (b) lattice site energies $V_n$ (real and imaginary parts) as given by Eq.(7) for
$N=15$, $\alpha=1+0.2i$ and $\gamma=-1+0.2i$. (c)
Numerically-computed temporal evolution of the site occupation
probabilities $|\psi(n,t)|^2$ in the $(n,t)$ plane for an input
Gaussian wave occupation amplitudes $\psi(n,0)=\exp\{-[(n-N/4)/4]^2
\}$. (d) Detailed temporal evolution of the occupation probability
$|\psi(0,t)|^2$ of the left-edge ($n=0$) lattice site. (e) Behavior
of total occupation probability $P(t)=\sum_n |\psi(n,t)|^2$.}
\end{figure}
with $M=0,1,2,...,N$. Owing to the equal spacing of energy
eigenvalues, the lattice model (7) yields a periodic temporal
dynamics with period $T=2 \pi/|\alpha-\gamma|$. Note that, for
$\alpha=-\gamma$ real-valued, this lattice model corresponds to the
one previously introduced in Refs.\cite{Lambropoulos,Gordon}, i.e.
to uniform site energies $V_n$ and to an optimal coupling of
adjacent lattice sites. Interestingly, for real values of $\alpha$
and $\gamma$ but with $\alpha \neq -\gamma$ and $\alpha \gamma <0$,
a linear gradient of site energies $V_n$ occurs, and the lattice
model (7) thus sustains exact BOs on a finite lattice. Figure 1
shows, as an example, a typical behavior of hopping amplitudes and
site energies for such an Hermitian lattice, together with the
characteristic BOs behavior for an initial Gaussian wave packet as
obtained by numerical simulations of the Schr\"{o}dinger equation $i
(\partial \psi/ \partial t)=\mathcal{H} \psi$. To show that exact wave packet reconstruction is insensitive to edge effects,
in the example of Fig.1 wave packet excitation was intentionally chosen close to one edge of the lattice, however
periodic wave packet reconstruction is observed for {\em any} initial wave packet, regardless of its shape or position in the lattice.
It should be also noticed that periodic wave packet reconstruction does not mean nor imply shape-invariance of the wave packet evolution.\\
Remarkably, a periodic quantum evolution, associated to the
existence of equally-spaced real-valued energies, is found even for
non-Hermitian lattices by taking $\alpha=\sigma+i \rho$ and
$\gamma=-\sigma+i \rho$, where $\rho$ and $\sigma$ are arbitrary
real-valued and nonvanishing constants. In this case, according to
Eq.(7) the hopping amplitudes $\kappa_n$ turn out to be real-valued
[$\kappa_n=\sqrt{(\rho^2+\sigma^2)(N-n)(n+1)}$], whereas the site
energies $V_n$ are complex-valued and linearly varying with the
index $n$ [$V_n=(\sigma+i \rho)N-2i\rho n$], i.e. the linear
gradient of site energies is now {\em imaginary} rather than
real-valued as in ordinary BO lattice models. Physically, such a new
class of non-Hermitian lattices could be realized by a sequence of
$(N+1)$ active optical waveguides with a judicious engineering of
waveguide spacing (to tune the hopping rates $\kappa_n$) and with a
controlled gain/loss coefficient $V_n$ (see, for instance,
\cite{Ruter10}). In spite of non-Hermiticity, the energy levels are
real-valued and equally spaced like in the Hermitian lattice model.
However, the kind of periodic wave packet evolution in the such a
non-Hermitian lattice is very distinct from ordinary BOs found in
Hermitian lattices (see e.g. \cite{BO2}). In particular,
oscillations of the total occupation probability $P(t)=\sum_n
|\psi(n,t)|^2$ are found as a result of the non-conservation of the
norm, like in other non-Hermitian crystals (see for instance
\cite{Christodoulides08}). As an example, Fig.2 shows the behavior
of $\kappa_n$ and $V_n$ for a non-Hermitian lattice defined by
Eq.(7), together with a typical temporal evolution of an initial
Gaussian wave packet, showing
a characteristic oscillation of the total occupation probability.\\
As a final comment, it is worth mentioning that BOs on a
semi-infinite Hermitian lattice with linearly-increasing hoping
amplitudes, recently predicted in Ref.\cite{Longhi09}, can be found
by assuming $F_1(n)=\alpha (n+1)$ and $F_2(n)=\gamma n$, with
$\alpha \gamma>0$ (to ensure lattice Hermiticity) and $|\alpha| \leq
|\gamma|$ (to ensure boundness of $\psi_0(n)$ as $n \rightarrow +
\infty$ \cite{note1}). In this case one obtains $\kappa_n= J (n+1)$
and $V_n=-fn-\alpha$ ($n=0,1,2,3,...$), where $J=\sqrt{\alpha
\gamma}$ and $f=(\alpha+\gamma)$ are the hopping rate and site
energy gradients, respectively. Note that the condition $|f| \geq
2J$ is always satisfied, i.e. the existence of a Wannier-Stark
energy level spectrum requires a minimum value of the dc force
$|f|$. As discussed in \cite{Longhi09}, such a minimum value
$|f|_m=2J$ of the
dc force corresponds to the existence of a metal-insulator transition in this lattice model.\\
\\
\begin{figure}
\includegraphics[width=78mm]{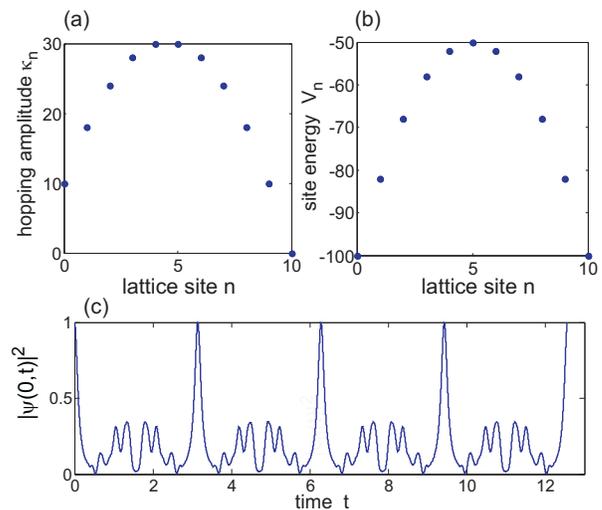}\\
   \caption{(color online) Self-imaging in an Hermitian lattice with commensurate energy levels that depend quadratically on the quantum
   number [Eqs.(9) and
   (10)]. Parameter values are $N=-\alpha=10$ and $\gamma=0$. (a) Behavior of  $\kappa_n$. (b) Behavior of $V_n$. (c)
Numerically-computed temporal evolution of the site occupation
probability $|\psi(0,t)|^2$ of the left-edge ($n=0$) lattice site
for initial single-site excitation $\psi(n,0)=\delta_{n,0}$.}
\end{figure}
{\em Self-imaging in lattices with commensurate
 energy levels quadratically-varying with the quantum number.} As a second class of lattices
showing self-imaging phenomena, let us consider a finite lattice
comprising the sites $n=0,1,2,...,N$ and assume
$F_1(n)=(n-N)(n+\alpha)$ and $F_2(n)=n(n+\gamma)$, where $\alpha$
and $\gamma$ are real-valued parameters. The hopping amplitudes and
site energies of this lattice are thus given by [see Eq.(2)]
\begin{eqnarray}
\kappa_n & = & \sqrt{(n+\alpha) (n-N)(n+1)(n+\gamma+1)} \;\;\;\;\;\; \\
V_n & = & -[2n^2+(\alpha+\gamma-N)n-\alpha N].
\end{eqnarray}
Substitution of such expressions of $F_1(n)$ and $F_2(n)$ into
Eq.(6) and by comparing the coefficient of the highest degree term
($\sim n^M$) in the power expansion, one obtains the eigenvalues
\begin{equation}
E=M^2-M(N+1+\gamma-\alpha), \;\;\;\;\; M=0,1,2,...,N,
\end{equation}
i.e. the energy eigenvalues are described by a quadratic function of
the quantum number $M$. Provided that $(N+1+\gamma-\alpha)$ is a
rational number, i.e. $(N+1+\gamma-\alpha)=r_1/r_2$ with $r_1$ and
$r_2$ integer numbers, the energy levels are commensurate numbers
and the wave packet dynamics turns out to be periodic. As an
example, Figs.3(a) and (b) show a typical behavior of $\kappa_n$ and
$V_n$ in an Hermitian lattice for parameter values $N=10$,
$\alpha=-N$ and $\gamma=0$, which yield $E=M^2-21M$ according to
Eq.(11). Periodic quantum evolution is thus expected at the period
$T=\pi$. Self-imaging is demonstrated, as an example, in Fig.3(c),
where the behavior of the site occupation probability
$|\psi(n=0,t)|^2$ at the lattice edge $n=0$ versus time $t$ for
initial single-site
excitation $\psi(n,0)=\delta_{n,0}$ is depicted.\\
\\{\em Conclusions.} A novel class of truncated
tight-binding lattices  that sustain a periodic quantum evolution of
the wave packet has been presented. In particular, we have shown the
existence of Bloch oscillations in Hermitian truncated lattices,
Bloch oscillations in non-Hermitian lattices with an imaginary site
energy gradient, and discussed self-imaging phenomena in finite
lattices with commensurate energy levels that depend quadratically
on the quantum number. Photonic waveguide arrays, realized for
instance by femtosecond laser writing in fused silica \cite{DL11},
could provide a possible experimental set-up to realize the
Hermitian lattice models discussed in this paper (Figs.1 and 3). In
such arrayed structures, simultaneous engineering of coupling rates
$\kappa_n$ and site energies $V_n$ can be achieved by appropriate
design of distances between adjacent waveguides (which determine the
values of $\kappa_n$) and choice of waveguide writing speed (which
determines the values of $V_n$).\\
\\
The author acknowledges financial support by the Italian MIUR (Grant
No. PRIN-2008-YCAAK).


\begin{thebibliography}{31}
\bibitem{note1}  For a
semi-infinite lattice, where $n=0,1,2,3,....$, $E=0$ is an
eigenvalue of $\mathcal{H}$ provided that $\psi_0(n)$ is bounded as
$n \rightarrow \infty$, which implies a constrain on the asymptotic
behavior of $F_1(n)$ and $F_2(n)$ as $n \rightarrow \infty$.
Conversely, for a finite lattice with $F_1(N)=0$ and $F_2(0)=0$,
$E=0$ is always an eigenvalue of $\mathcal{H}$.
\end{thebibliography}

\end{document}